\documentclass[a4paper]{jpconf}
\usepackage{graphicx}
\usepackage[utf8]{inputenc}
\usepackage{amsmath,amssymb}
\usepackage{hyperref}
\usepackage{listings}
\usepackage[russian,english]{babel}
\usepackage{color}
\usepackage[colorinlistoftodos]{todonotes}
\usepackage[numbers,sort&compress]{natbib}

\definecolor{litered}{RGB}{255,112,112}

\definecolor{dkgreen}{rgb}{0,0.6,0}
\definecolor{gray}{rgb}{0.5,0.5,0.5}
\definecolor{mauve}{rgb}{0.58,0,0.8}

\begin{document}
\title{Toolbox for multiloop Feynman diagrams calculations using $R^{*}$ operation}

\author{D.V.~Batkovich
, M.V.~Kompaniets}
\address{Department of Theoretical Physics, St.~Petersburg State University, Uljanovskaja 1,
St. Petersburg, Petrodvorez, 198504 Russia}
\ead{batya239@gmail.com, mkompan@gmail.com}

\begin{abstract}
We present the toolbox for analytical calculation of $UV$-counterterm of Feynman diagrams. It combines the power of $R^{*\prime}$-operation with modern analytical methods. Written in pure \texttt{Python} our toolbox can be easily used and extended.
\end{abstract}

\section{Introduction}

In this paper we present the toolbox that allows one to calculate $UV$-counterterm of Feynman diagrams in the framework of Bogoliubov–Parasiuk $R'$-operation~\cite{bog1957} within $MS$-scheme (see e.g.~\cite{bogshirk, vasilev}). It uses a number of methods that include infrared rearrangement  ($IRR$)~\cite{vladik1978, chet6}, $R^{*}$-operation~\cite{chet, chet2, chet4, chet5}, $G$-functions method~\cite{chet6} and integration-by-parts ($IBP$) reduction~\cite{baikov2010, lee, lee2, lee2012, lee2012dra, chet3, lee4, baikov}.

Infrared rearrangement is one of the oldest tricks for calculating $UV$-counterterm of diagrams, it is based on the fact that $UV$-counterterm of a logarithmically divergent diagram is independent on external momenta and masses of this diagram~\cite{collins}. This fact allows one to set some masses or external momenta to zero or to introduce new masses or external momenta, the only restriction is that such a rearrangement should not produce nonphysical infrared (IR) divergences. It is possible to overcome this restriction by $R^{*\prime}$-operation and significantly increase the number of counterterms that can be calculated using $G$-functions and $IBP$-reduction. 

Toolbox presented in this paper  can be applied to a wide range of models. Additionally, it can be  easily  extended to calculate graphs not only using $G$-functions and $IBP$-reduction, but also with some other analytical or numerical methods (graph calculators) using API (application program interface) provided.

The paper is organized as follows: in the next section we shortly discuss $R^*$-, $R^{*\prime}$- operations, $IR$-counterterms and present an example that shows benefits of $R^{*\prime}$-operation. The third section describes architecture of the toolbox and packages it contains. In the fourth section we present examples on how to calculate $UV$-counterterms using  \texttt{RStar} package. Fifth section is devoted to the implementation of the $IBP$-reduction and the API for the graph calculator extensions. The last section shortly describes toolbox installation procedure.


\section{Short introduction to $R^{*\prime}$ operation}
$R^{*}$-operation was introduced in~\cite{chet4, chet5} as a method that allows to overcome restrictions of the infrared rearrangement: if one calculates $UV$-countertem of the diagram with improper $IRR$ using $R^{\prime}$-operation there are some nonphysical $IR$-divergences arising in the integrals calculated. This $IR$-divergences as well as $UV$-divergences are poles in $\varepsilon$ (parameter of the dimensional regularization) and can not be distinguished from $UV$-divergences, this results in the incorrect $UV$-counterterm. $R^{*\prime}$-operation allows to separate this nonphysical $IR$-divergences and subtract them to get correct value for counterterm. Thus counterterm $\Delta_{UV}(\gamma)$ for diagram $\gamma$ is defined as a combination of standard $R^{\prime}$ operation and infrared $\widetilde R^{\prime}$ operation:
\begin{equation}
\Delta_{UV}(\gamma)=-KR^{*\prime}(\gamma)=-K\widetilde{R}
^{\prime}R^{\prime}(\gamma)\end{equation}
In the case of the proper infrared rearrangement, i.e. no IR divergences, $\widetilde R^{\prime}\equiv 1$, but in the case of improper IRR it becomes nontrivial.
$\widetilde{R}'$-operation can be defined by introducing $IR$-counterterm ($\Delta_{IR}$) as follows:


\begin{equation}
\widetilde{R}'(\gamma)=\left\{ \begin{array}{ll}\displaystyle
\Delta_{IR}(\gamma) & \textrm{, where $\gamma$ is $0$-tadpole,}\\
\\
\displaystyle
\gamma + \sum\limits_{\delta \in W_{u}(\gamma)}\Delta_{IR}(\gamma/\delta)\delta & \textrm{, other,}\\
\end{array} \right.
\end{equation}
here $W_{u}(\gamma)$ - set of all $UV$-divergent subgraphs of $\gamma$ through which external momentum of $\gamma$ can be completely passed (these subgraphs contain all boundary vertices of initial graph) and which also don't contain tadpoles. By $0$-tadpole we denote massless tadpole. 
If one treats 0-tadpole as integral it is definitely $0$, but in case 
of $R^{*\prime}$-operation combinatorics it is important to take these graphs into account. $IR$-counterterm in case of $0$-tadpole $\gamma$ can be given by the following relation:
\begin{equation}
\Delta_{UV}(\gamma) + \Delta_{IR}R(\gamma) = 0.
\end{equation}
Advantages of using $R^{*\prime}$-operation can be demonstrated by 6-loop diagram example:
\vskip -7mm
\begin{equation*}
\begin{matrix}
\includegraphics[width=5cm]{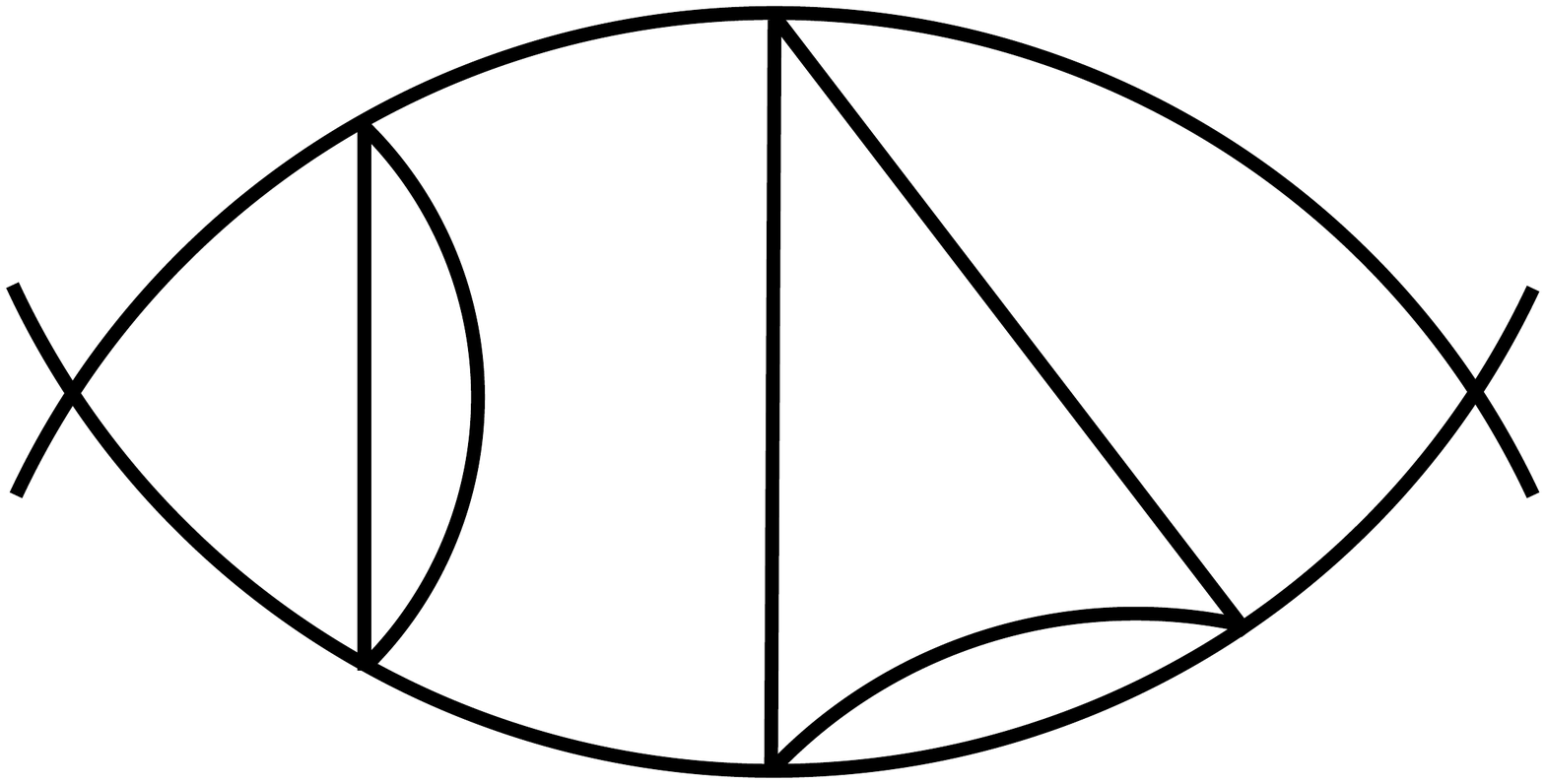} 
\end{matrix}
\quad
\begin{matrix}
\includegraphics[width=5cm]{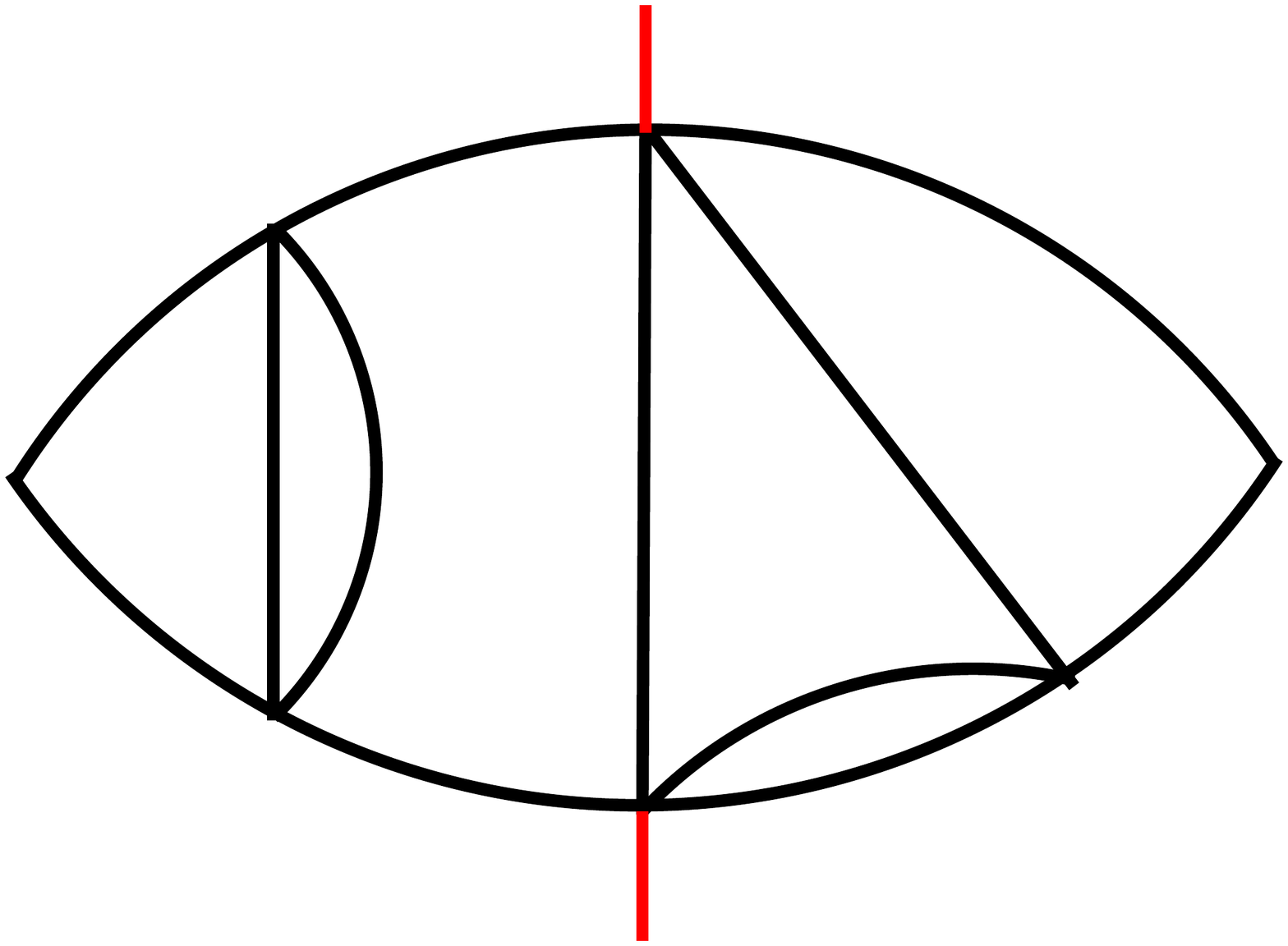} 
\end{matrix}
\end{equation*}
\vskip -7mm
$UV$-counterterm of the left diagram can't be calculated with proper $IRR$ by using $G$-functions an 4-loop $IBP$-reduction, but if we rearrange external momenta(and set all masses to zero) as shown on the right diagram $UV$-counterterm can be calculated by using $R^{*\prime}$ with use of $G$-functions method only.

Further information about $R^{*}$- and $R^{*\prime}$-operations can be found in~\cite{chet, chet2, chet4, chet5}. 

\section{Architectural overview}

Toolbox is built on top of \texttt{GraphState} and \texttt{Graphine}~\cite{graphstate} libraries, all graph manipulations and serialization are provided by these libraries. All input graphs must be given in form readable by  \texttt{GraphState}~\cite{graphstate} which is the extension of graph labeling algorithm introduced by~\cite{nick}.

\texttt{RStar} package plays the central role in the toolbox and is responsible for all combinatorial operations to evaluate $R^{*\prime}$-operation. Additionally this module can calculate components of $R^{*\prime}$, namely $\Delta_{IR}$ and $\widetilde{R}$ of some given graph.

\vskip -9mm
\begin{equation*}
\includegraphics[width=4.8cm]{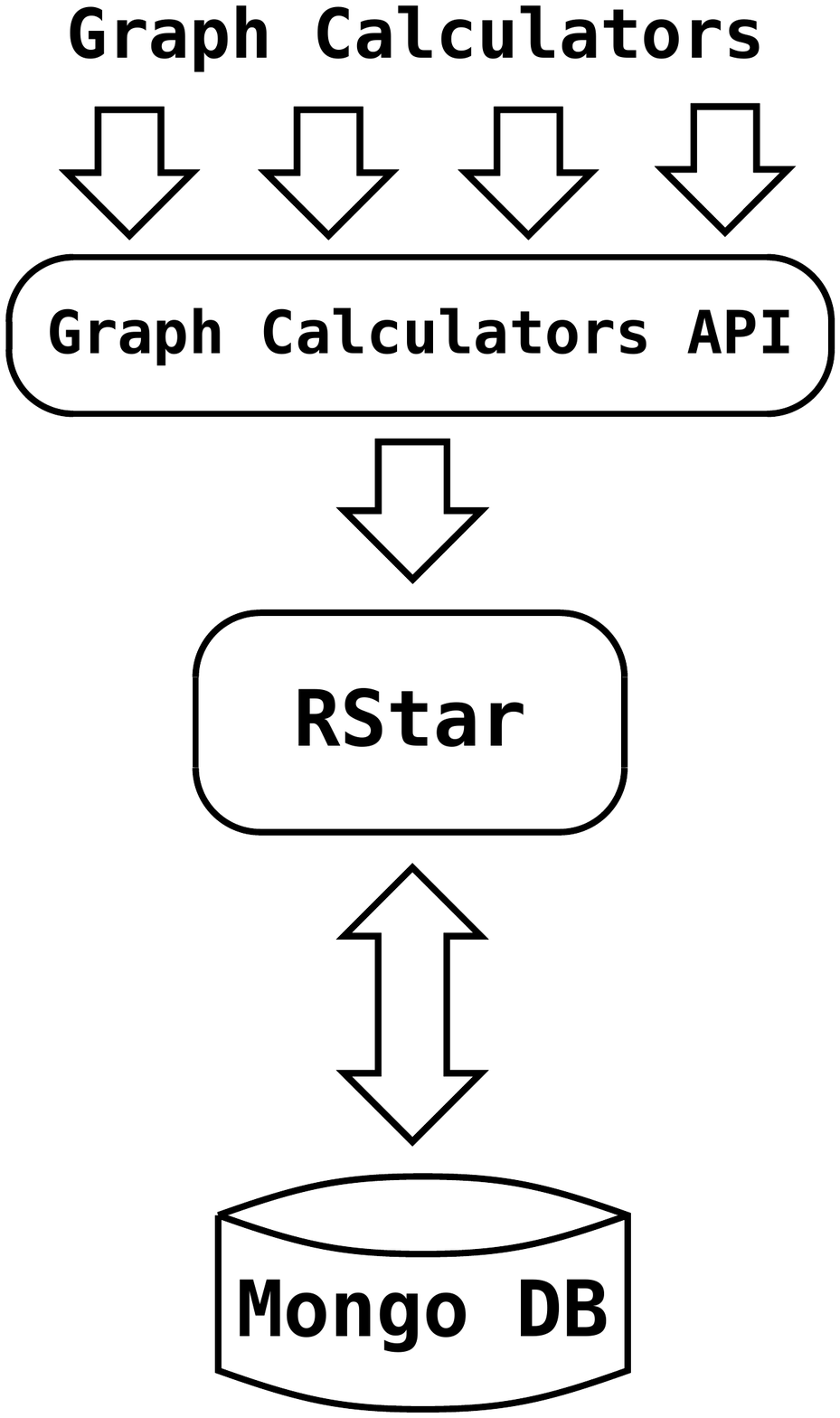}
\end{equation*}
\vskip -25mm
\subsection{Brief packages overview}
\begin{itemize}
\item\texttt{RStar} - main package serves to calculate $\Delta_{UV}$, $\Delta_{IR}$ and $\widetilde{R}'$ on diagrams. Package contains graph calculators API to calculate required diagram values. Default list of calculators can be extended. Calculated values of diagrams and counterterms are placed on \texttt{Mongo DB}~\cite{mongo} -based storage. This storage saves all values which can be reused for another diagrams and has separate collections for different methods and models. Environment of this package can be simply configured as shown below.

\item\texttt{RgGraphEnv} and \texttt{RgGraphUtil} - some set of utils containing connector to \texttt{Mongo DB}, \texttt{swiginac} CAS extensions and $G$-functions calculator.

\item\texttt{Reduction} package contains implementation of $IBP$-reduction and wrappers of this procedure for graph calculators API.
\end{itemize}

\subsection{Symbolic calculations}

As backend for computer algebra we use \texttt{swiginac} library. It's a python wrapper of the most known \texttt{GiNaC} computer algebra written in \texttt{C++}. It gives us fast analytical computations and provides better performance than pure \texttt{Python} libraries. 


\section{Getting started with \texttt{RStar}}

Different models require different environmental parameters. We provide a simple API to configure user own environment including user own graph calculators, $UV$-relevance conditions to determine $UV$-divergent subgraphs (note that condition of 1-irreducibility is automatically applied), dimensions as well as \texttt{Mongo DB} storage settings. These parameters must be configured before starting any calculations as shown in the following program code example: 

\begin{lstlisting}
from rstar import Configure
from rggraphenv.symbolic_functions import cln, e
from rggraphenv import GLoopCalculator, StorageSettings
from graphine import Graph
from graphine.filters import graph_filter

d_phi4 = cln(4) - cln(2) * e

@graph_filter
def uv_condition(graph_edges, super_graph):
    g = Graph(graph_edges)
    uv_index = 4 * g.loops_count - 2 * g.internal_edges_count
    return uv_index >= 0

Configure()\
  .with_uv_filter(uv_condition)\
  .with_dimension(d_phi4)\
  .with_calculators(GLoopCalculator(d_phi4))\
  .with_storage_holder(StorageSettings("test_project", "localhost", 27017))\
  .configure()
\end{lstlisting}

When the program environment is configured one can use $R^{*\prime}-$operation to evaluate $UV$- and  $IR$-counteritems. All the calculated data will be automatically saved to structured storage. In the following snippet we show how to calculate $\Delta_{IR}$ and $\Delta_{UV}$ on examples of:

\begin{figure}[h!]
\begin{equation*}
\Delta_{IR}\Big (
\begin{matrix}
\includegraphics[width=3cm]{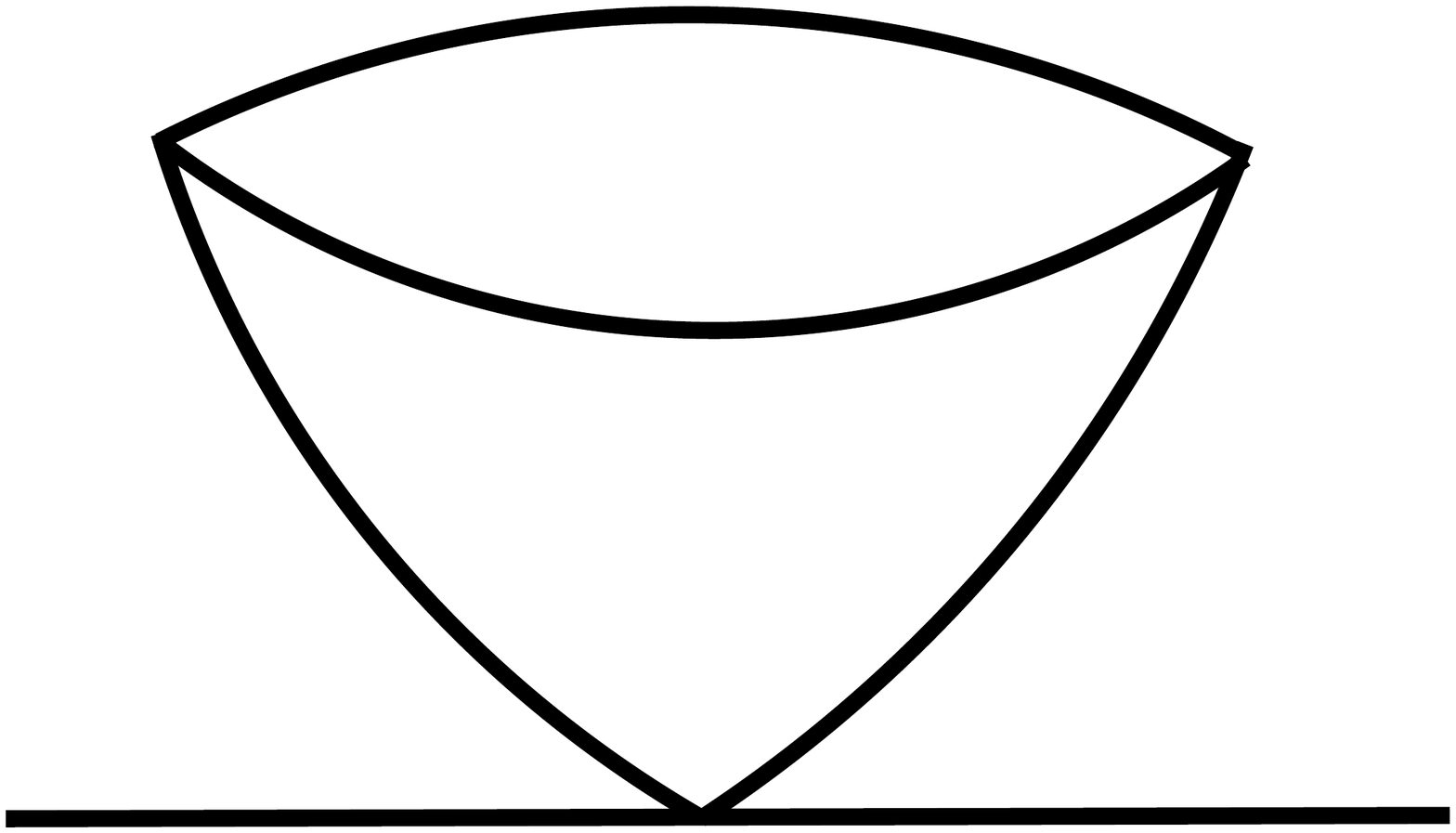} 
\end{matrix}
\Big)
\quad
\text{and}
\quad
\Delta_{UV}\Big (
\begin{matrix}
\includegraphics[width=3cm]{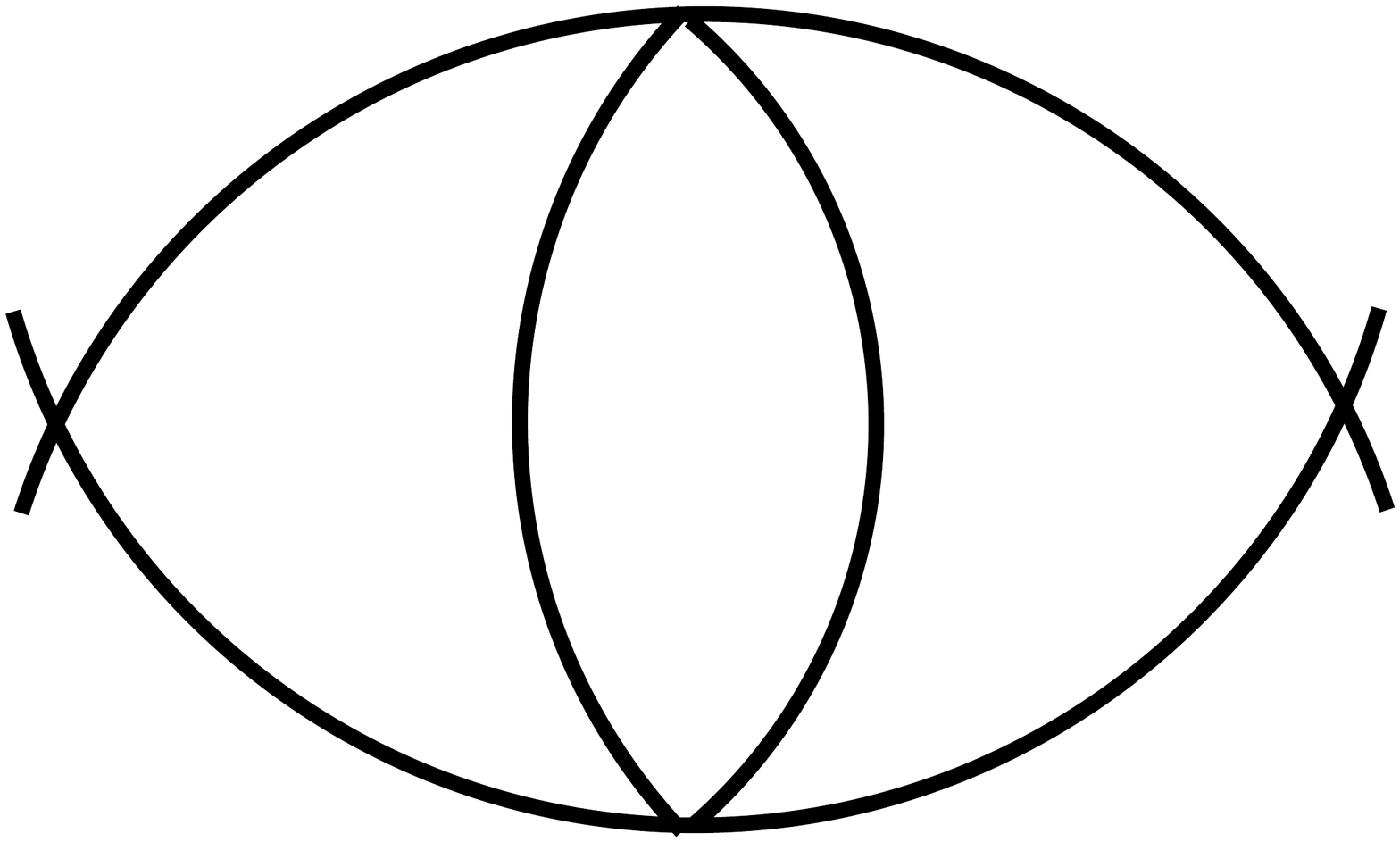} 
\end{matrix}
\Big)
\end{equation*}
\caption{Diagram examples.}
\label{fig:diags}
\end{figure}

To calculate these counterterms user needs to create \texttt{RStar} object and call correspondent methods (\texttt{delta\_ir} and \texttt{delta\_uv}):
\begin{lstlisting}
from rstar import RStar, graph_util
operator = RStar()

eye_tadpole = graph_util.graph_from_str("ee12|22||")
print operator.delta_ir(eye_tadpole)
# 1/2*(1+e)*e**(-2)

three_loop_diagram = graph_util.graph_from_str("ee12|223|3|ee|")
print operator.delta_uv(three_loop_diagram)
# -1/3*e**(-1)+2/3*e**(-2)-1/3*e**(-3)
\end{lstlisting}

Note that both of graphs must be given in \texttt{GraphState}~\cite{graphstate}  graph representation based on Nickel index~\cite{nick, graphstate}. In current example $ee12|22||$ is representation of left diagram and $ee12|223|3|ee|$ is representation of right diagram pictured on Fig.~\ref{fig:diags}.

\section{Calculator extensions}

\texttt{RStar} graph calculator API allows to create custom calculators. These calculators can be written in any language and can use any software or hardware instruments. To create custom calculator one should implement \texttt{GraphCalculator} class

\begin{lstlisting}
from rggraphenv.abstract_graph_calculator import AbstractGraphCalculator

class MyCustomGraphCalculator(AbstractGraphCalculator):
    def get_label(self):
        return "description of calculator"
        
    def init(self):
    	# do initialization here
        pass
        
    def dispose(self):
        # clean resources here
        pass
        
    def is_applicable(self, graph):
        # check is calculator applicable to graph
        return can_calculate_graph

    def calculate(self, graph):
        # calculation logic...
        return calculated_value
\end{lstlisting}

In the \texttt{calculate} method the logic of the user graph calculation process  must be incapsulated. Return value must be \texttt{None} if the calculator can't calculate a given graph or a pair where first component is value of \texttt{swiginac} expression type that depends on $\varepsilon$.
Second component must have the type \texttt{rggraphutil.VariableAwareNumber} and is representing power of $p^{-2}$ where $p$ is the external momentum of the graph:
\begin{lstlisting}
import rggraphutil
rggraphitul.VariableAwareNumber("l", a, b)
\end{lstlisting}
means $a+bl$ number where $l=d/2-1$ and will represent $p^{-2(a+bl)}$. For example:
\begin{lstlisting}
def calculate(self, graph):
    # calculation logic...
    return 1/rggraphenv.symbolic_functions.e, VariableAwareNumber("l", 1, -1)
\end{lstlisting}
means that returned value equals to $\frac{1}{\varepsilon}p^{-2(1-l)}$.

\subsection{{\normalfont\texttt{Reduction}} package}

\texttt{Reduction} provides reduction by $IBP$~\cite{chet3} and $DRR$~\cite{lee2012dra, tarasov2012, baikov} rules and implemented using rules of {\color{litered}\bf LiteRed}~\cite{lee, lee2, lee4} program. Values of master integrals for $4$-loop reduction obtained from~\cite{baikov2010, lee2012}. In our implementation we use some cache for more effective utilization of already calculated integrals. Our cache is common for all run reduction processes at the same time and stored to \texttt{Mongo DB}. Additionally cache is separate for different reduction rules such as rules for 2, 3 and 4 loop reduction. \texttt{Reduction} can be used separately from other parts of the program.

To be calculated Feynman integral must be given in the following form:
\begin{equation}
J(n_1, n_2, n_3,\ldots, n_m) = \int \frac{\prod\limits_l^{L} dk_{l}}{D_1^{n_1}D_2^{n_2}\ldots D_m^{n_m}},
\end{equation}
where $D_{i}$ - i-th propagator of L-loop basis. To define this integral from \texttt{Python} one need following code:

\begin{lstlisting}
import reduction
integral = reduction.J(n1, n2, ..., n_m)
\end{lstlisting}

One can see the list of basis propagators by typing \texttt{reduction.BASIS\_2}, \texttt{reduction.BASIS\_3} or \texttt{reduction.BASIS\_4}. Last number means count of loops in basis.

But sometimes it can be very uncomfortable to generate these \texttt{reduction.J} objects manually from diagrams. For this case we allow to set graph as the parameter and our package will find corresponding momentum layout. The only thing which are needed to specify in this case is how to obtain propagator weights. This procedure is carried out using additional function that has one parameter (edge) and returns weight of propagator corresponding to given edge. It must be specified by user.

Here is the example how to calculate following diagram with weights of propagators shown in picture: 
\vskip -9mm
\begin{equation*}
\includegraphics[width=5cm]{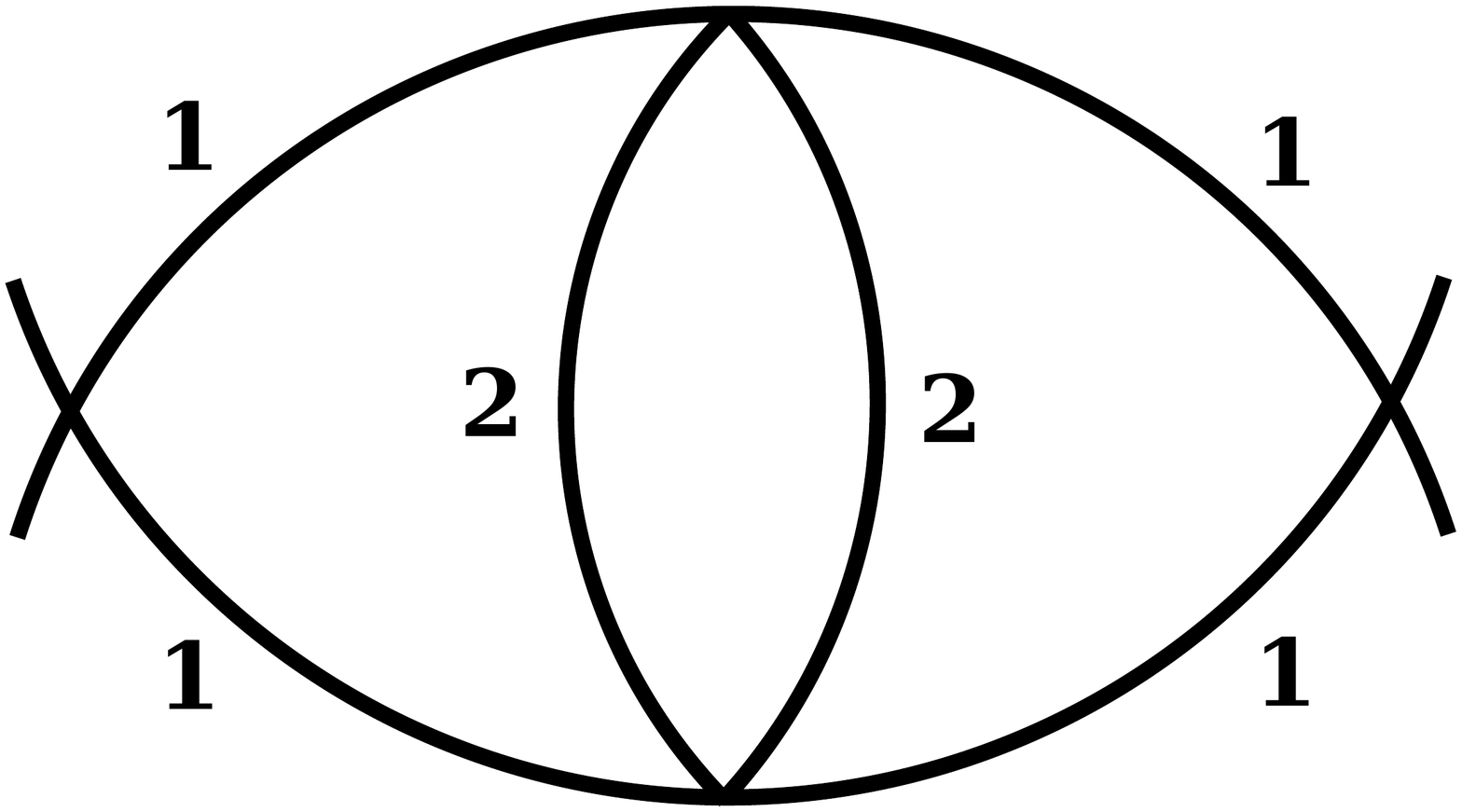}
\end{equation*}
\vskip -6mm
To calculate this diagram we use graph representation and integral representation. For graph representation we define function \texttt{weight\_extractor} that returns \texttt{weight} attribute of given edge.

\begin{lstlisting}
import reduction
from rggraphenv import symbolic_functions

reductor = reduction.THREE_LOOP_REDUCTOR.with_cache("localhost", 27017)

weight_extractor = lambda e: e.weight

def calculate_3loop_graph_using_reduction(graph):
    unsubstituted_result = reductor.calculate_diagram(graph, weight_extractor)
    # in variable unsubstituted_result:
    #     - master integral values not substituted
    #     - dimension not substituted
    #     - no expansion in series by epsilon
    return unsubstituted_result.evaluate(substitute_masters=True,
                                             _d=4-2*symbolic_functions.e,
                                             series_n=4,

                                             remove_o=True)


def calculate_3loop_integral(integral):
    unsubstituted_result = reductor.calculate_j(integral)
    return unsubstituted_result.evaluate(substitute_masters=True,
                                             _d=4-2*symbolic_functions.e,
                                             series_n=4,
                                             remove_o=True)


def main():
    three_loop_diagram = reduction.graph_util.graph_from_str("e12|223|3|e|:0_1_1|2_2_1|1|0|")
    print calculate_3loop_graph_using_reduction(three_loop_diagram)
    # 1/3*e**(-3)+1/3*e**(-2)+1/3*e**(-1)+...

    three_loop_diagram_as_integral = reduction.J(1, 1, 0, 2, 2, 0, 1, 0, 1)
    print calculate_3loop_integral(three_loop_diagram_as_integral)

if __name__ == "__main__":
    main()                     
\end{lstlisting}
Result of \texttt{calculate\_j} and \texttt{calculate\_diagram} methods is linear combination of master integrals, where values of these master integrals and dimensions in coefficients are not substituted. To substitute them  we call \texttt{evaluate} method. It should be noted that in bases included in \texttt{Reduction} package values of master integrals are given only for dimension $d=4-2\varepsilon$. If one need to perform calculations in dimension that differs from $d=4-2\varepsilon$, it is necessary to clone basis needed and set values of master integrals for required space dimension. Also it is possible to create own basis and reduction rules. 

\section{Installation}

Our toolbox supports only Unix-like systems. As external dependencies one need \texttt{swiginac}, \texttt{pymongo}, \texttt{repoze.lru} and \texttt{inject} packages which one can find in \texttt{Python} \texttt{Package}  \texttt{Index} \url{https://pypi.python.org/}. Additionally following packages are needed to install \texttt{GraphState}, \texttt{Graphine}, \texttt{RgGraphEnv}, \texttt{RgGraphUtil}, \texttt{Reduction}, \texttt{RStar} from \url{https://code.google.com/p/rg-graph/}.

\section{Acknowledgments}

Authors are grateful to L.Ts.~Adzhemyan, K.G.~Chetyrkin and R.N.~Lee for helpful discussions and advices, to ACAT’14 Organizing Committee for support and hospitality. The work was supported in part by Saint-Petersburg State University (project 11.38.185.2014). 

\bibliography{paper}

\end{document}